\documentclass[a4paper,10pt]{article}
\pdfoutput=1 
\usepackage{jcappub} 
\usepackage[T1]{fontenc} 
\usepackage{graphicx}
\usepackage{subfig}
\usepackage{hyperref}
\usepackage{amstext,amsmath,amssymb,amsfonts,bbm, amsthm}
\def\app#1{{Appendix~\ref{#1}}}
\def\sec#1{{Section~\ref{#1}}}
\def\eqs#1{{Eqs.~(\ref{#1})}}
\def\fig#1{{Fig.~\ref{#1}}}

\def\be{\begin{equation}}
\def\ee{\end{equation}}
\def\bes{\begin{eqnarray}}
\def\ees{\end{eqnarray}}
\def\nn{\nonumber}

\title{\bf Constraints on Lepton Asymmetry from Nucleosynthesis in a Linearly Coasting Cosmology}

\author[a]{Parminder Singh}
\author[a,b]{Daksh Lohiya}

\affiliation[a]{Department of Physics and Astrophysics, University of Delhi, Delhi 110007, INDIA.}
\affiliation[b]{DAMTP, Centre for Mathematical Sciences, Cambridge, CB3 OWA, UK}

\emailAdd{$^a$psingh@physics.du.ac.in}
\emailAdd{\\~~~~~~~~~~~$^b$dl116@cam.ac.uk}

\abstract{
We study the effect of neutrino degeneracy on primordial nucleosynthesis in a universe in which the cosmological scale factor evolves linearly with time. The degeneracy parameter of electron type neutrinos ($\xi_e$) determines the $n/p$ (neutron to proton) ratio, which in turn determines the abundance of $^4$He in a manner quite distinct from the Standard Scenario.
The observed abundances of $^4$He, $\mathrm{Y}_P$=0.254$\pm$0.003, and the minimum metallicity that is essential for fragmentation and cooling processes in star forming prestellar gas clouds (Z = Z$_{cr}$ = 10$^{-6}$Z$_\odot$), constrain the baryon to photon ratio, $\eta_B$=(3.927$\pm$0.292)10$^{-9}$, corresponding to a baryonic matter density, 
$\Omega_B$=0.263$\pm$ 0.026 and $\xi_e$=-2.165$\pm$0.171. This closes the dynamic mass estimates of matter in the universe, obtained from large scale velocity dispersion in galaxy clusters, by baryons alone. Useful byproducts are the threshold X(CNO) abundances required to trigger the CNO cycle in the observed low metallicity stars in the universe.}
\begin{document}
\maketitle
\flushbottom

\section{Introduction}
\label{sec:intro}

A universe with an assumed large scale homogeneity and isotropy is described by the FRW metric:
\begin{equation}
ds^2 = c^2dt^2 - a^2(t)\left\{\frac{dr^2}{1 - kr^2} + r^2(d\theta^2
+ sin^2\theta d\phi^2)\right\}.
\end{equation}
The scale factor $a(t)$ is determined by requiring this metric to be a solution 
to the Einstein equations for a given equation of state for matter. The 
``$\Lambda_{\mathrm{CDM}}$'' cosmological model, preceded by a suitable inflationary 
epoch, defines what is now referred to as the ``Standard Scenario'' of cosmology. 
The parameters of the model are determined by constraints from a wide range of cosmological observations. Obtaining consistent constraints of the model parameters in the Standard Scenario is a very impressive accomplishment. However, this consistency has two ``weakest links'': (i) Dark matter, which, to date, has escaped direct detection by current Particle Physics experiments and (ii) Dark energy, or the Cosmological Constant($\Lambda$), whose small value poses a major theoretical challenge.\\

A crucial cornerstone of the Standard Scenario is the Standard Big Bang 
Nucleosynthesis (SBBN). The SBBN predicted light element abundances are fairly 
concordant with observations. However, there have been reservations on the 
compatibility of observed estimates of $^4$He and $^7$Li with numerical estimates 
in the standard scenario \cite{steg07,cburt08}. Attempts to resolve this problem by invoking the 
effect of a probable non-vanishing neutrino degeneracy parameter have been made with limited 
success \cite{steg08,bg1976,steg04}. Adjusting the neutrino degeneracy parameter to get the right amount of Helium becomes incompatible with Lithium abundance measurements. \\

The SBBN also has issues with missing metallicity [Z] which is defined by the 
abundance of all elements heavier than $^7$Li\cite{alain,iocco}.  Light elements 
produced in the nucleosynthesis epoch come with an abysmally low collateral 
production of metallicity (Z$\sim\mathcal{O}(10^{-16}$)). Any metallicity enrichment 
in the standard scenario could only be expected after the formation and 
disintegration of very high mass PopIII stars that may produce a critical metallicity 
(Z$_{cr}\sim10^{-6}Z_{\odot}$). This minimum Z = Z$_{cr}$ is essential for an 
efficient cooling process and fragmentation of star forming prestellar gas clouds that 
lead to the formation of the later generation of low metallicity stars
\cite{sch2006,omu2005,escan}.
The required exploding PopIII stars should be observed as magnitude 27 to 29 stars
at any time in every square arc-minute of the sky. However, no such sighting has 
been reported to date - in spite of advances in observational sensitivity 
 to detect such dim objects \cite{jmesc}. \\

The above issues, amongst other reasons, have motivated us to look at alternative 
cosmological models. Of particular interest is the ``Power Law Cosmology'' in which 
the cosmological scale factor evolves as $a(t)\propto t^\alpha$, with $\alpha \sim 1$.
Such an evolution accommodates high red-shift objects, alleviating the ``age problem'', 
it is purged of the fine tuning issues, and is a generic feature in a class of models 
that attempt to dynamically solve the cosmological constant problem 
\cite{adev1,dol,wein,ford,allen,mann}. 
Such an evolution on the large scale could emerge as a feature of a cosmology in 
finite range ``massive gravity'' theories that are defined on a background 
``reference'' flat Minkowski spacetime. These theories are purged of the pathology 
of the original attempts of formulating a theory of a low mass graviton by Fierz and 
Pauli \cite{fierz}. An open FRW metric with the scale factor evolving linearly with time 
also arises in the fourth order ``Weyl'' gravity model \cite{weyl1,weyl2}. \\

The motivation for such an endeavor has been further discussed in a series 
of earlier articles \cite{adev1,adev,mset,gset1}.
Dolgov and Ford demonstrated that a non-minimally coupled scalar field rapidly 
develops a scalar condensate whose stress tensor rapidly cancells any large 
positive cosmological 
constant in the theory. Concurrently, the scale factor $a(t)$ rapidly 
approaches a linear evolution. Unfortunately the effective gravitational 
constant becomes vanishing small. However, a similar cancellation of the 
cosmological constant, without the cancellation of the effective gravitational
constant, can be realized in theories having free massless vector 
and tensor fields minimally coupled to gravity \cite{dol1}. The back reaction of the 
energy momentum tensor of these fields again exactly cancels a background 
cosmological constant. This cancellation implies and comes with the power law evolution of 
the scale factor. The magnitude of the stress tensor of normal matter fields 
being always much smaller than the cosmological constant, these normal fields have negligible 
effect on the scale factor evolution. However, perturbations of the normal matter fields 
around a large scale homogeneous and isotropic distribution, satisfies 
perturbed Einstein equations. A similar behaviour of the scale factor is seen in 
the fourth order conformally invariant model of 
Mannheim and Kazanas \cite{mann}. In these models the variation of the action, 
having a Weyl invariant 
scalar part, with respect to the metric tensor, gives a rank two tensor that 
identically vanishes for the
conformally flat FRW metric. A non-minimally coupled scalar field then produces
an effective \emph{repulsive gravitation} that results in the FRW scale factor
quickly approaching $a(t) \longrightarrow t$. Non-conformally flat 
perturbations around this linear
coasting background is again effectively described by perturbed Einstein equations. \\
\\
A linear evolution of the scale factor also follows from 
the fact that the use of Einstein's equations as such in cosmology has
never ever been justified. The averaging problem and the continuum limit
in General Relativity have not been properly addressed to. In any case, most
treatments have not considered the retarded effects in their full 
generality\cite{ellis,ehlers,tavak}. Averaging Einsteins 
equations across horizon lengths lacks self consistency. On the other hand,  
Newtonian cosmology, applied to an exploding {\it Milne ball}
in a flat space-time [see eg. 
\cite{milne,rindler}] gives a unique linear coasting cosmology viz. the FRW 
[{\it{Milne}}] metric with $a(t) = t$.
Such a cosmology does not suffer from the horizon problem. 
A linearly evolving model has neither a particle horizon nor a cosmological 
event horizon.
Linear evolution that is independent of the matter equation of state is 
also purged of the flatness or the fine tuning problem
\cite{adev1,dol,ford,adev,mset,gset1} as the evolution of the scale factor in such theories
does not constrain the matter density parameter to be close to a critical 
value. \\

There are accounts of Einstein's theory of gravity arising as an effective 
theory of a spin two field in flat spacetime. Gravitation can be attributed to 
the interaction of a massless (spin-two) field with a source distribution.  
One starts with a source of gravitation having a compact support in a flat Minkowski spacetime 
and determines its interaction with 
a spin two field. Iteratively, one considers the stress tensor of the induced gravitational 
field in every order of perturbation itself as an additional source. This 
leads one to recover Einstein's theory \cite{Feynman}. As a matter of fact, 
just a few (one or two) iterations brings the effective theory sufficiently close 
to, and experimentally indistinguishable from, Einstein's theory. The induced 
spin two field is determined by the convolution of its Green's function 
(propagator) over the moments of the source distribution. We may take this 
iterative prescription as defining gravitational interactions 
for any distribution. It is safe to conjecture that as a distribution satisfying the 
cosmological principle has vanishing moments of the source
distribution, it gives a vanishing correction at the very first order. The
iterative correction therefore vanishes for all orders. Thus we would expect 
a distribution satisfying the Cosmological principle to coast freely - as if 
there was no gravitation driving the background
whatsoever. Thus Einstein's theory would hold only for local perturbations around the 
homogeneous and isotropic distribution of matter.\\

We may take any of the above as the basis for our linear coasting conjecture. 
In what follows, we assume that a homogeneous background FRW universe
is born and evolves as a Milne Universe and perturbations around this 
background are described by perturbed Einstein equations.  In this article we 
concentrate of nucleosynthesis in such a model. \\

At the very outset, we make no claim of having yet found a definitive 
viable linear coasting  
alternative cosmological model or theory. Indeed the amount of effort that has 
gone to establish 
concordance of cosmological observations with the standard $\Lambda_{CDM}$ 
prior is quite commendable. The purpose is to outline aspects of nucleosynthesis in
linear coasting models in their own right. Many of the theoretically interesting
alternative models described above, that successfully addressed to deep 
cosmological conundrums, were abandoned rather prematurely 
just because their 
characteristic power law evolution for the cosmological scale factor being  
quite different from that in SBB - they were wrongly regarded as
unsustainable without any further ado. \\
\\
For the purpose of this article, we shall conjecture that a universe with a matter distribution obeying the cosmological principle (namely: possessing large scale homogeneity and isotropy), is described {\it{in the large scale}} by the FRW metric with:
\begin{equation}
a(t) \propto t
\end{equation}
independent of the equation of state of matter.
We shall refer to such a universe as a linearly coasting universe. Such an evolution 
of the cosmological scale factor, is a surprisingly nice fit to a host of cosmological observations\cite{adev1,adev,mset,gset1,chardin,rgv1,rgv2}. In this article, we revisit aspects of primordial nucleosynthesis in such a linearly coasting universe following from a straightforward modification of standard nucleosynthesis codes \cite{wag69,wag73,kaw88,kaw92}.\\

A linearly evolving scale factor leads to an early universe (linearly coasting) nucleosynthesis (LCN) quite distinct from SBBN. This is outlined briefly in \sec{lcn}. It turns out that the predicted relic Cosmic Neutrino Background(CNB) in LCN is quite similar
to that in the Standard Model(SM) with an effective neutrino temperature T$_{0\nu}$ $\sim$ 1.94 K. This relic CNB may possibly hide a large asymmetry in the number density of neutrinos and their antineutrinos due to non vanishing
degeneracy parameters $\xi_\alpha$, ($\alpha=e,\mu,\tau$), corresponding to the three neutrino flavours. These parameters are assumed to vanish in {\it{minimal}} SBBN. However of late it has been suggested that the observed primordial $^4$He abundance is inconsistent with SBBN with a vanishing neutrino degeneracy parameter.\\

A non-vanishing neutrino degeneracy parameter is an essential feature in models in which baryons and charged leptons are created as a result of a phase transition in a spontaneously broken gauge theory. Such a transition requires the net lepton number after symmetry breaking, to be of the same order as the net photon number in the universe [see \cite{linde} for a review]. Achieving this by a non-vanishing $\xi_\alpha$ that may give a net relic lepton number density in the form of neutrinoes as high as $2.5 n_\gamma$, is not ruled out by present cosmological constraints.\\

As stated before, SBBN with $\xi_\alpha\ne0$ can accommodate the observed primordial $^4$He abundance, however this leads to inconsistency with the observationally inferred primordial abundances of $^7$Li and the minimum metallicity requirements \cite{steg07,steg08,bg1976,steg04,alain,iocco}.\\

In LCN, a non zero $\xi_e$ crucially affects initial neutron to proton ratio and its evolution. Previous studies of LCN neglected the electron neutrino degeneracy($\xi_e$) and
reported rather high estimates of baryonic density required for the production of observed
amounts of $^4$He \cite{mset}. Unfortunately, such a high baryonic density over - closes
dynamic mass estimates of matter in the universe obtained from large scale velocity dispersion in galaxy clusters, roughly by a factor of three. $^4$He abundance is sensitively dependent on the baryon density as well as the neutron to proton ratio. With a non vanishing $\xi_e$, the required baryonic density, necessary for the observed $^4$He, can be brought down to acceptable levels, and gives 
rise to yields of minimum metallicity as a useful by product.\\

\sec{lcn} outlines the hot universe at the time of nucleosynthesis. A linearly evolving scale factor leads to a drastic reduction of the expansion rate of the universe in comparison to that in SBBN at the same temperature. This extends time scales at high temperatures and results in a significantly large metallicity production. In \sec{LAAL}, we incorporate the effect of electron neutrino degeneracy ($\xi_e$) and the associated ``Neutron to Proton ratio'' in LCN. \sec{NDPA} describes constraints on $\{\Omega_B,\xi_e\}$ for the production of the critical metallicity and the observed $^4$He abundances.

\section{Nucleosynthesis in a Linearly Coasting Universe}
\label{lcn}

With $a(t) \sim t$, the Hubble parameter is simply:
\begin{equation}
 H(t)\equiv \frac{\dot{a}(t)}{a(t)}= \frac{1}{t}
\end{equation}
Measurement of the current value of the Hubble parameter: $H_0 = 73.8\pm2.4$ km/sec/Mpc\cite{rie2011}, gives an estimate of the   present age of the universe as some 13$\times$10$^9$ years. The age of the universe at the epoch of primordial nucleosynthesis, when
 temperatures are of the order of T$\sim$ 10$^9$ K and higher, follows from the present effective CMB temperature, $T_0$= 2.73 K, and the relation between the  scale factor of the universe and temperature:
\be
a(t)T(F(T))^{\frac{1}{3}} = constant = a_0T_0(F(T_0))^{\frac{1}{3}}
\ee
\be
\label{RT}
\Rightarrow t=\frac{t_0T_0F(T_0)^{\frac{1}{3}}}{T(F(T))^{\frac{1}{3}}}
\ee
These expressions follow from entropy conservation in terms of the density and 
pressure of light particles. $F(T)$ is given by [see eg. \cite{weinberg}]
\begin{equation}
\label{FT}
F(T) =1+\frac{\rho_{e^-}+\rho_{e^+}+p_{e^-}+p_{e^+}}{\rho_\gamma+p_\gamma}
\end{equation}
It varies continuously from 2.75 at high temperatures much greater than $10^{10}$K, 
to unity at low temperatures much less than $10^9$K. This reduces \eqs{RT} to
\be
t=\frac{T_0}{H_0T(F(T))^{\frac{1}{3}}}
\ee
This gives the age of universe to be some 36 years at T$\sim$ 10$^9$ K. Thus such an evolution envisages an expansion much slower in comparison to that, at corresponding temperatures, in the Standard Model (SM).\\

The entropy of neutrinos does not significantly affect the above result. When the temperature is greater than a few Mev, the energy density of the universe is dominantly due to  photons, electron-positron pairs, neutrinos ($\nu_\alpha$) and antineutrinos ($\bar\nu_\alpha$), $(\alpha=e,\mu,\tau)$, along with a contamination of baryons, in thermal equilibrium. At such high temperatures and densities, (neutral current) weak interactions:
\begin{equation*}
e^+ + e^- \leftrightarrow \nu_\alpha + \bar\nu_\alpha
\end{equation*}
are rapid enough to maintain neutrinos in thermal equilibrium with the
photon-e$^\pm$ plasma. With the universe expanding very slowly, leptonic weak interactions do not decouple till  temperatures as low as $\sim 10^8$ K. However these reactions not decoupling is necessary but not sufficient to keep the neutrinoes in thermal equilibrium with photons. As the temperature drops to values lower than $\sim 5\times10^9$ K, the e$^\pm$ pairs start getting  annihilated into photons, with the (reverse) pair creation of e$^\pm$ suppressed due to lesser number of high energy photons. This results in an increase
of photon temperature in comparison to the $\nu - \bar\nu$ temperature. The neutrinos and antineutrinos would not get
heated up by this e$^\pm$ annihilation due to the small branching ratio of the weak interaction channel
in comparison with electromagnetic annihilation channel:
$(e^+ +e^- \rightarrow \gamma +\gamma)$.
After the e$^\pm$ annihilation, the photon temperature $(T_\gamma)$ gets enhanced by a factor of $(11/4)^{1/3}$ in comparison with
the neutrino effective temperature $(T_\nu)$ - just as in the SM. This would lead to a relic
Cosmic Neutrino Background(CNB) at present with an effective temperature, 
T$_{0\nu} \sim 1.94$ K.\\

At high temperatures, the neutron to proton density ratio is determined by equilibrium sustained by their forward and backward weak interaction rates with leptons. It is roughly given by:
\begin{equation}
\label{np}
\left(\frac{n}{p}\right)_{eq} \simeq \exp \left(\frac{-15.0}{T_9} -\xi_e \right)
\end{equation}
With a(t)$\sim$t, the slow expansion rate of universe ensures that electron neutrino weak interactions with nucleons do not freeze out even till temperatures as low as $\sim 10^9$ K. In particular the inverse beta decay, which converts protons into neutrons, does not freeze out at these temperatures. This keeps replenishing more neutrons into nucleosynthesis network \cite{mset}. Thus even though the $n/p$ ratio keeps decreasing below temperatures $10^{10}$K, the induction of neutrons by the inverse beta decay of the proton is a characteristic  feature of LCN that enables the production of $^4$He up to the desired observed levels.
A modification of the standard nucleosynthesis code incorporating these peculiarities of LCN is quite straightforward. High temperatures that are sustained for  large enough times due to a slower evolution, allow for successive burning of
helium, carbon and oxygen, leading to significant  metallicity production in LCN roughly some
10$^8$ times the metallicity produced in the SBBN. These enhanced levels of metallicity produced in LCN are close to the minimum metallicity required for the cooling and fragmentation of collapsing proto - stellar gas clouds as suggested by
\cite{omu2005}. A by product is X(CNO) abundances, at sufficient levels to initiate and sustain the CNO cycle in the old low metallicity stars of the universe \cite{cno}.\\

Unfortunately, the slow expansion in LCN leads to abysmally low levels of residual deuterium(D) and lithium($^7$Li). However, it has been pointed out that spallation mechanism in incipient PopII star environments can produce acceptable levels of D and $^7$Li in the later history of the universe \cite{gset}. Spallation as a mechanism for deuterium production was reviewed by Epstein et al \cite{eps}, who considered spallation of a beam of ions, with the constitution and abandance of nucleii of elements observed in typical incipient type I stars, on a target cloud - again of the same constitution. There was no problem of producing Deuterium to desired levels. However, producing Deuterium up to such desired levels over produced $^7$Li. On the other hand, constraining $^7$Li to the observed levels produced virtually no Deuterium. Revisiting the above analysis, one finds that $^7$Li is mainly produced by $\alpha$ in the beam spalling over $\alpha$ in the target cloud, and, to a much smaller extent, by protons in the beam spalling over heavier nucleii in the target cloud. This is smaller due to the smaller density of heavier nucleii. The same spallation mechanism was reviewed in \cite{gset} in incipient type II stellar environments. Stellar flares in a protostar provide a natural source of beams of ions that would be deficient in alpha particles which would spall over metal deficient clouds to give deuterium to observed levels without overproducing $^7$Li in the later history of the Universe. We return to this issue in Section 5. 
\\

At the nucleosynthesis epoch in LCN, for $\xi_e = 0$, successful production of desirable 
$^4$He and metallicity levels put constraints on the baryonic mass density 
$\Omega_B \sim 0.70$. This over-closes dynamic mass estimates obtained from observed 
velocity dispersion of stars in galaxies and clusters\cite{om}. 
There are at least two ways of dealing with this 
problem. \\

Firstly: assuming the universe to have witnessed a first order quark-gluon 
phase transition, the picture that would emerge is that of bubbles of hadron 
phase produced inside a quark plasma. The nucleation of such bubbles 
that would precede the transition results in a net baryon number 
distrubuted at the walls of colliding bubbles of true, hadronic ground 
state. This would result in a very inhomogeneous distribution of hadrons 
within an almost uniform radiation background. At very high temperatures 
above $\approx 10^{10}$K, charged current interactions would keep all 
baryons strongly coupled to the radiation background. For large enough 
bubbles, this would prevent the baryons distribution from homogenizing. 
However, at lower temperatures, once the charged current interactions tend 
to freeze, neutrons would start diffusing while the protons would remain  
locked with the radiation. This would give rise to a sharp variation of
the neutron proton ratio and, in turn, the helium production - particularly
in the linear slow evolution where the amount of helium is very sensitive 
to the $\eta$. A simple minded two zone calculation can proceed as follows. 
Considering a fraction {\it{``r''}} of a given co-moving volume having 
an effective baryon density parameter $\Omega_B^{(1)}$, and the remaining 
fraction $1 - r$ having  an effective density parameter $\Omega_B^{(2)}$. 
Choosing  $\Omega_B^{(1)}$ to be large enough to produce 100\% Helium, and 
 $\Omega_B^{(2)}$ small enough to give almost no Helium, it is straight 
forward to see that the overall  $\Omega_B$ and the net $Y_{He}$ are given 
by
\begin{equation}
 Y_{He}\Omega_B = r \Omega_B^{(1)}
\end{equation} 
We can have a whole range of  $\Omega_B^{(1)}$ to give the required $Y_{He}$ 
by keeping  $\Omega_B$ within the dynamic mass estimate bounds.\\

Secondly, again assuming the universe to have emerged out of a first order
QGP phase transition, it has been demonstrated that in a large number of
scenarios with a net baryon number, a net neutrino lepton number density 
$|n_\nu - n_{\bar{\nu}}|$ of the order of magnitude of the photon number 
density is expected an essential byproduct \cite{linde}. 
Describing this in terms
of a non-zero neutrino degeneracy parameter $\xi$, the $n/p$ ratio is  
significantly affected in terms of eq(8).
A combination of both the above effects, that are in turn 
related to the physics of the hadronic phase transition, can in principle
yield concordance with nucleosynthesis and dynamic mass bounds which can 
be closed with baryons alone.\\

In the next section we see how one can secure concordance of the closure of dynamic 
mass with baryons alone, just with an appropriate
non vanishing neutrino degeneracy parameter $\xi_\alpha$. \\

\section{Lepton Asymmetry and the Neutron to Proton Ratio}
\label{LAAL}
\subsection{Lepton Asymmetry}
\label{LA}

The number density n$_\nu$, and mass density $\rho$,  of neutrinos of either kind, is determined by the degeneracy parameter  $\xi_\alpha$:
\begin{equation}
n_{\nu_\alpha,\bar\nu_\alpha} \equiv \frac{(kT_\nu)^3}{2\pi^2(\hbar c)^3}
\int_{0}^{\infty}\frac{x^2\mathrm{d}x}{1+\exp(x\mp\xi_\alpha)}
\label{nnu}
\end{equation}

\begin{equation}
\rho_{\nu_\alpha,\bar\nu_\alpha} \equiv \frac{(kT_\nu)^4}{2\pi^2c^2(\hbar c)^3}
\int_{0}^{\infty}\frac{x^3\mathrm{d}x}{1+\exp(x\mp\xi_\alpha)}
\label{rhonu}
\end{equation}
the $\mp$ being respectively for neutrino and antineutrinos respectively,
and $\alpha (\alpha=e,\mu,\tau)$ representing its flavour. T$_\nu$ is the neutrino temperature. $\xi_{\alpha}$ is related to chemical potential as
$\xi_{\alpha}\equiv \mu_\alpha/kT_\nu$.
A lepton asymmetry parameter that determines the asymmetry between neutrinos and
antineutrinos can be defined as:
\begin{equation}
\eta_L \equiv \sum_\alpha L_\alpha
\end{equation}
\begin{equation}
L_\alpha=\left(\frac{n_{\nu_\alpha}-n_{\bar\nu_\alpha}}{n_\gamma}\right)
\end{equation}
Where, $n_\gamma$ is number density of photons:
\begin{equation}
n_\gamma = \frac{2\zeta(3)}{\pi^2}\left(\frac{kT_\gamma}{\hbar c}\right)^3
\end{equation}
This gives:
\begin{equation}
\label{etal}
\eta_L\equiv \frac{1}{4\zeta(3)}\left(\frac{T_\nu}{T_\gamma}\right)^3
\sum_\alpha f(\xi_\alpha)
\end{equation}
\begin{equation}
\label{le}
L_e\equiv \frac{1}{4\zeta(3)}\left(\frac{T_\nu}{T_\gamma}\right)^3
 f(\xi_e)
\end{equation}
with  $\zeta(3)\simeq$ 1.202, and 
\begin{equation}
f(\xi_{\alpha})\equiv\left(\int_{0}^{\infty}\frac{x^2\mathrm{d}x}{1+\exp(x-\xi_{\alpha})}-
\int_{0}^{\infty}\frac{x^2\mathrm{d}x}{1+\exp(x+\xi_{\alpha})}\right)
\end{equation}
\begin{figure}[t!]
\includegraphics[width=0.45\textwidth,scale=0.5]{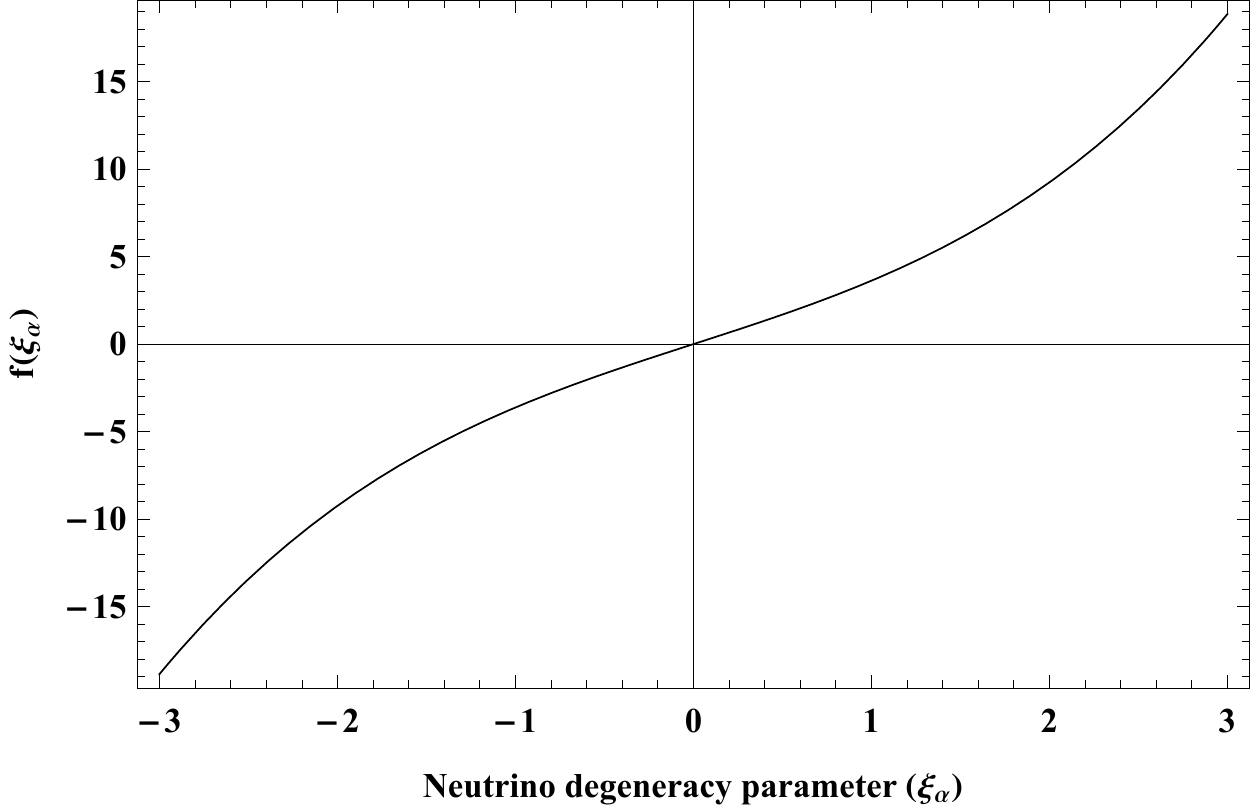}
\centering
\caption{Asymmetry function f($\xi_\alpha$) as a function of neutrino
degeneracy parameter $(\xi_\alpha)$}
\label{denfun}
\end{figure}
Here f($\xi_{\alpha}$) is the asymmetry function \fig{denfun}.   A non vanishing  $\xi_e$ yields a non zero $L_e$. This plays an important role in the charged current weak interactions that regulate the inter conversion of neutrons and  protons during the nucleosyntheis era through
the following reactions.
\bes
\label{reaction}
\mathrm{p + e^-\leftrightarrow n + \nu_e} \nn\\
\mathrm{n \leftrightarrow p + e^- + \bar\nu_e} \nn \\
\mathrm{n + e^+ \leftrightarrow p + \bar\nu_e}
\ees
The expression for production and destruction rates
of neutrons and proton are given in ~\app{apn}.
These reaction rates are plotted as a function of temperature (in units of T$_9 = 10^9$K) in \fig{forev} for different $\xi_e$ values.

\subsection{The Neutron to Proton Ratio}
\label{npratio}

The primordial abundance of $^4$He is quite sensitive to the neutron abundance during the nucleosynthesis epoch. A non zero $\xi_e$ would alter the number densities of $\nu_e, \bar\nu_e$, or equivalently L$_e$, which in turn modifies the rate of n$\leftrightarrow$p weak interactions and thereby altering the neutron to proton ratio.
\begin{figure}[t!]
\includegraphics[width=0.45\textwidth,scale=0.5]{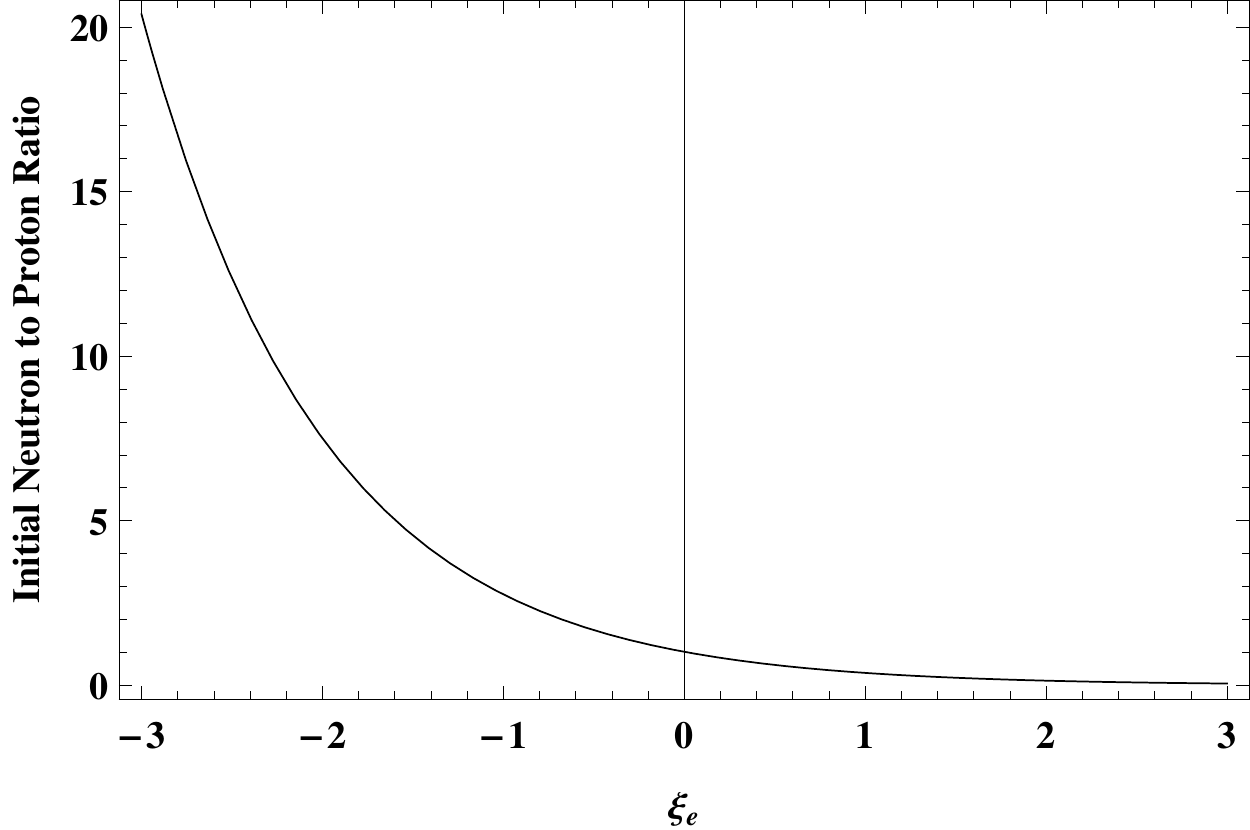}
\centering
\caption{Initial neutron to proton ratio as a function of electron neutrino
degeneracy parameter $(\xi_e)$}
\label{npxi}
\end{figure}
We trace the evolution of neutron to proton ratio ($n/p$) from very high temperatures T $\sim$ 10$^{11}$ K, when
 the n$\leftrightarrow$p rates are much faster than the expansion rate of the universe. We assume chemical equilibrium to have set in between neutrons and protons.
The initial $n/p$ ratio, $T \gtrsim 10^{11}$K is plotted as a function of $\xi_e$ in \fig{npxi}.
For $\xi_e<0, (L_e<0)$ one has an overabundance
of antineutrinos leading to a higher $n/p$ ratio while
 $\xi_e>0, (L_e>0)$ gives an overabundance of neutrinos, leading to a lower initial $n/p$ ratio.
The modification of charged-current weak interaction rates of the reactions given in \eqs{reaction} are
plotted as a function of temperature in units
10$^9$ K in \fig{forev}.
We compare the inverse beta decay rate with hubble expansion rate in \fig{hubrev}
for different values of $\xi_e$.\\

\begin{figure*}[t!]
\includegraphics[width=0.45\textwidth,scale=0.2]{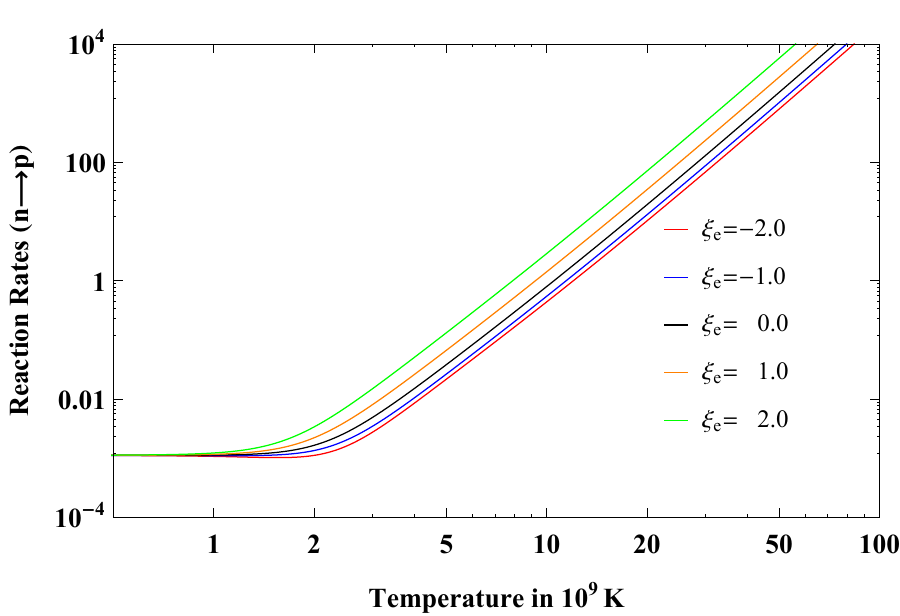}\hfill
\includegraphics[width=0.45\textwidth,scale=0.2]{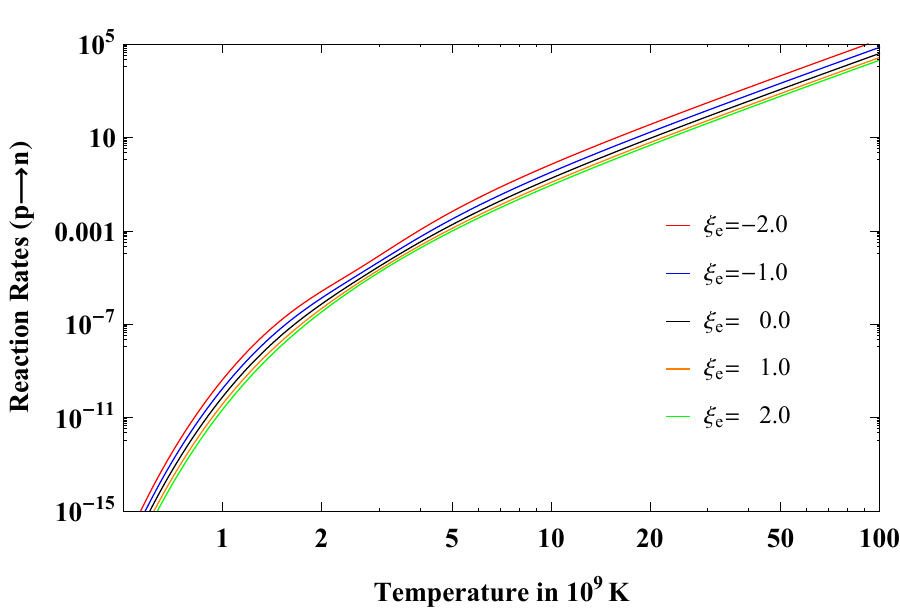}\hfill\\
\caption{Plot of the forward reaction rates $($n$\rightarrow$p$)$ and reverse reaction rates $(\mathrm{p}\rightarrow \mathrm{n})$ as a function of temperature in units 10$^9$ K for different values of electron neutrino degeneracy parameter $(\xi_e)$}
\label{forev}
\end{figure*}
\begin{figure*}[t!]
\centering
\includegraphics[width=0.49\textwidth,scale=0.2]{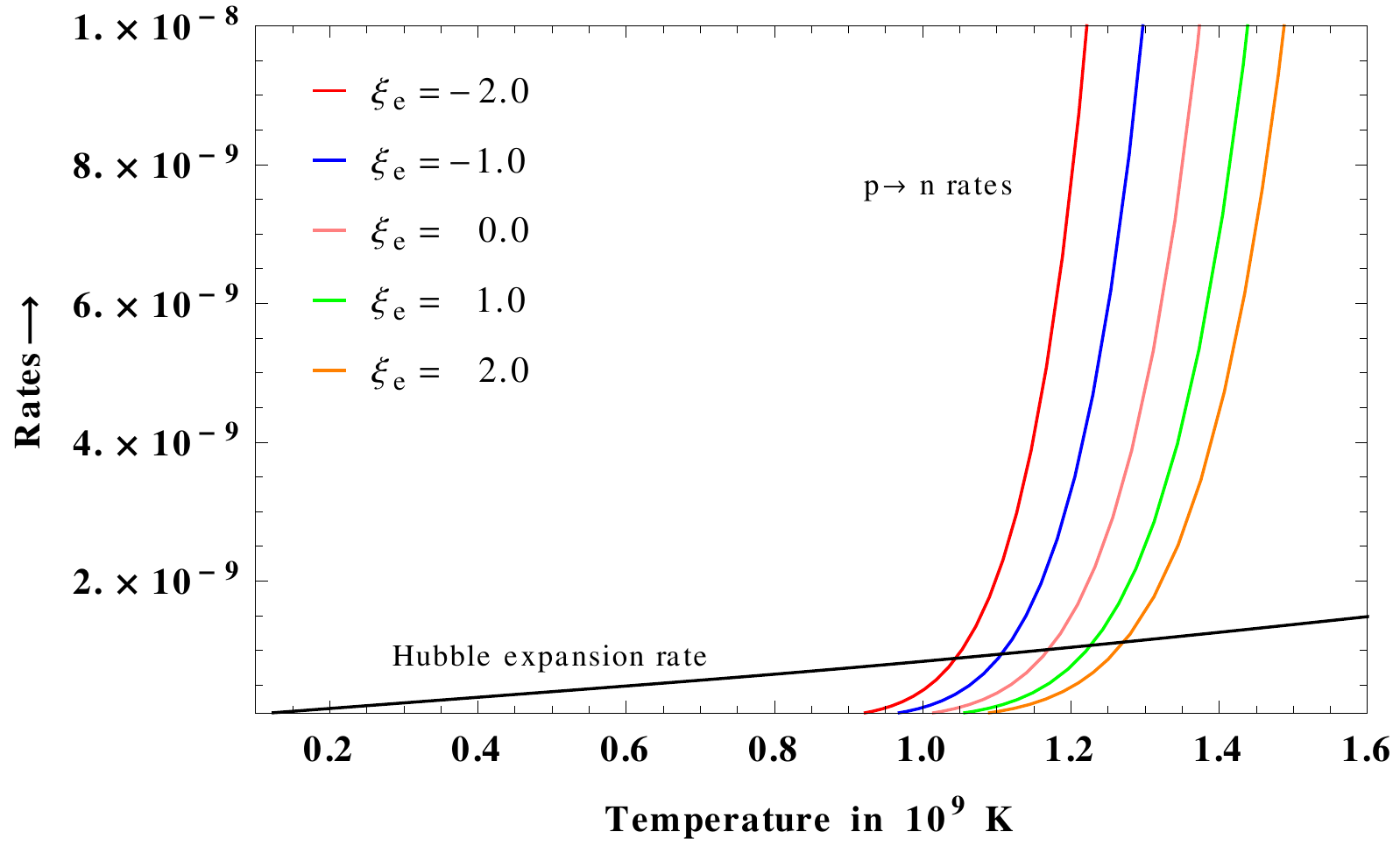}\hfill
\caption{ The inverse beta decay rate(p$\rightarrow$ n) for different $\xi_e$
values and Hubble expansion rate
as a function of temperature in units 10$^9$ K }
\label{hubrev}
\end{figure*}
\begin{figure*}[t!]

\includegraphics[width=0.47\textwidth,scale=0.2]{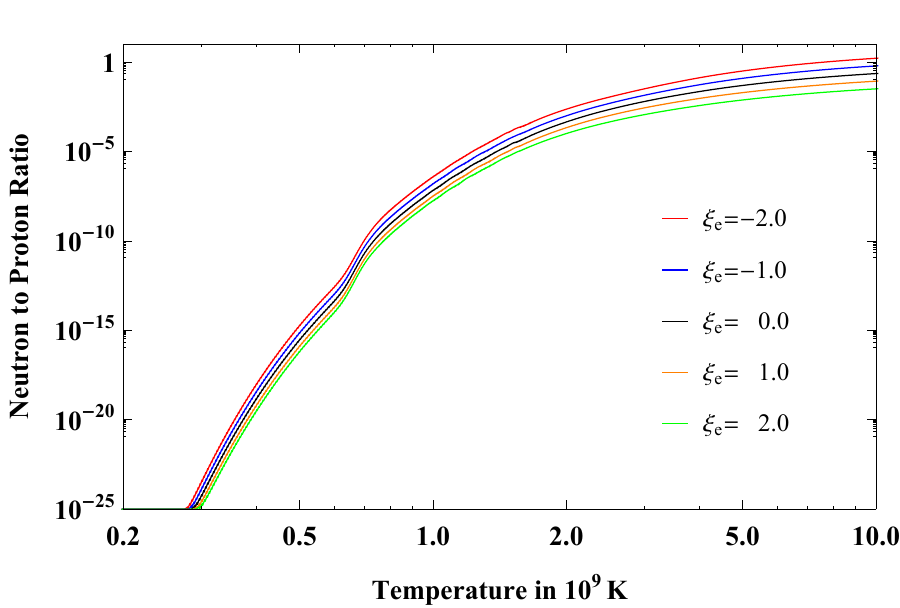}\hfill
\includegraphics[width=0.48\textwidth,scale=0.2]{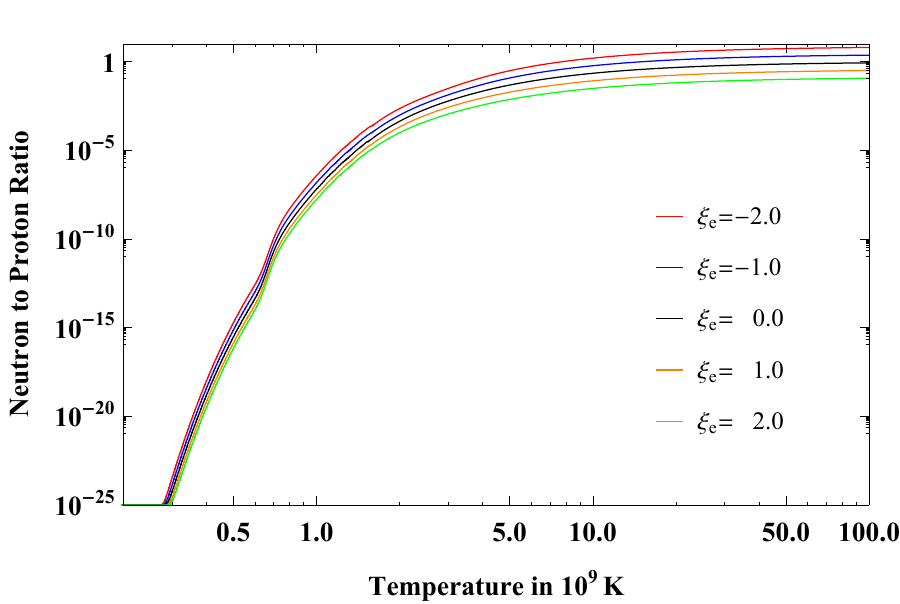}\hfill

\caption{Evolution of neutron to proton ratio as a function of temperature in units 10$^9$ K
for different $\xi_e$ values}
\label{fullrat}
\end{figure*}
For temperatures T $>10^{10}$ K, the n$\leftrightarrow $p
reactions are much faster than the hubble expansion rates.
This should ensure chemical equilibrium between the neutrons and protons. Before the commencement of nucleosynthesis, the $n/p$ exactly tracks its
equilibrium value. Significant nucleosynthesis
begins at T$\sim7\times10^{9}$ K.
Neutrons branch off to the light element formation network.
The slow
expansion rate of universe shifts the freezeout temperature of inverse beta decay to
very low values in comparison with BBN. This ensures replenishing of neutrons, due to inverse beta decay of the protons, to the equilibrium value and ensures significant nucleosynthesis.
For T$<10^9$K, the p$\rightarrow $n reactions
freezeout, but with n$\rightarrow $p reactions remaining active, $n/p$ gets depleted at very fast rates
\fig{fullrat}.
For $\xi_e$ positive and increasing, the rate of n$\rightarrow $p
reaction increases and p$\rightarrow $n decreases. This leads to significantly lower neutron levels in comparison with $\xi_e$=0 and results in a low $n/p$ at the time of nucleosynthesis. The freezeout temperature of p$\rightarrow $n shifts towards
higher values. For negative
values of $\xi_e$, the forward reaction decreases and the reverse reaction
gets enhanced. This enhances $n/p$ value, and at the same time, the freezeout temperature of p$\rightarrow$n
shifts toward lower values.\\

LCN and BBN have distinctive features.
In BBN a non zero $\xi_e$
affects primordial abundances primarily due to change in the expansion rates of the Universe. A
non zero neutrino degeneracy parameter always increases the
matter density of the universe and thereby the
expansion rate. This increment in expansion rate causes a freeze out of
weak interactions at higher temperatures resulting in an enhanced residual n/p ratio
at the nucleosynthesis epoch which changes very slowly after the freeze out due to the decay of neutrons. Thus significant nucleosynthesis occurs well after the n/p ratio is frozen. In LCN on the other hand, the hubble expansion rate is independent of density and is not affected by the increase of
neutrinos and antineutrinos by any non zero $\xi_e$. However, the n/p keeps to its equilibrium value that
depends on $\xi_e$ but not on $\xi_\mu$ and $\xi_\tau$. Weak interactions remain in equilibrium throughout the epoch when significant nucleosynthesis occurs. One can constrain $\xi_\mu$ and $\xi_\tau$ to be the same as $\xi_e$ (as done in 
\cite{steg08}), as it is now 
accepted, from flavour oscillation results, that the three neutrino distributions should have the same shape with a common value for $\xi$.  In the following we consider bounds on $\eta_L$ from such constraints. \\

\section{Neutrino Degeneracy and Predicted Abundances}
\label{NDPA}
LCN with a vanishing neutrino degeneracy requires a high $\eta_B$ in order to get right amount of $^4$He.
 \be
 \eta_B\simeq1.05\times10^{-8}
 \ee
 This parameter is related to $\Omega_B$ by:
\be
\eta_B=n_B/n_\gamma=\frac{273.9~\Omega_Bh^2}{10^{10}}
\ee
\be
\eta_9=\eta_B\times10^9=27.39~\Omega_Bh^2
\ee
$\Omega_B$ being the ratio of present baryon mass density and critical
density.
\be
\Omega_B\simeq0.70
\ee
Such high estimates of baryonic mass density overcloses dynamic mass estimates roughly
by a factor of three.\\

A non - vanishing $\xi_e$ significantly changes the equilibrium n/p ratio and thereby the production of $^4$He.
Estimates of the observed primordial abundance of $^4$He can therefore serve as a good ``leptometer''. The primordial metallicity(Z) being quite sensitive to $\eta_B$, can serve as a good ``baryometer''.
A straightforward modification of the standard NUC-123 (``Kawano'') code, incorporating the nuances of linear coasting,
is described in \cite{parm}.
$^4$He and metallicity(Z) levels produced are smooth and monotonic functions of $\eta_B$, $\xi_e$ and $H_0 = 100 h$ km/sec/Mpc. An empirical fit for Y$_P$ in the range of $\eta_9$ (0.3 $\lesssim\eta_9\lesssim 10$) is found to be:
\be
Y_P=0.2400\pm(0.0007)+0.3102(\eta_{Y}-0.7647h-1.8420)
\ee
Where,
\be
\label{etay}
\eta_Y=\ln(\eta_9) - 0.5009\xi_e\hspace{.8cm}
\ee
\begin{figure*}[t!]
\centering
\includegraphics[width=0.47\textwidth,scale=0.2]{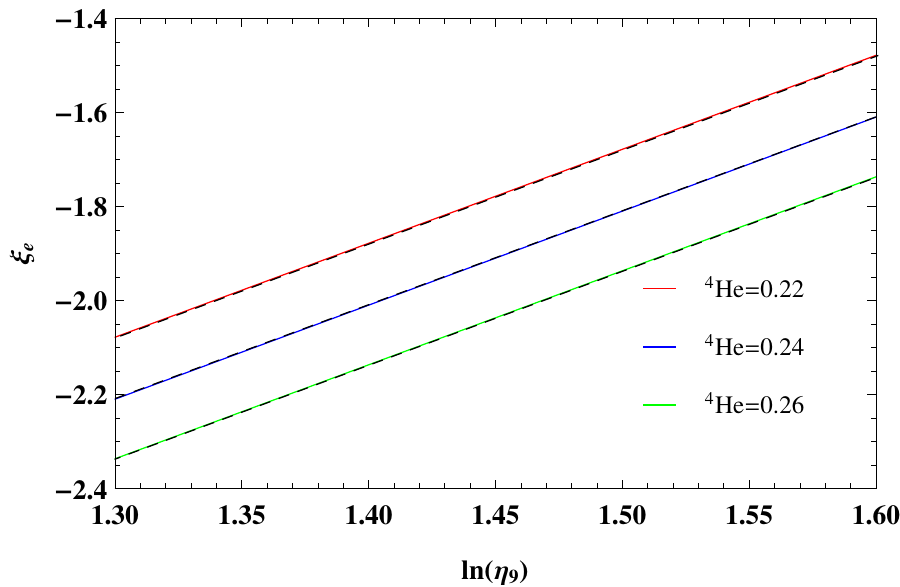}\hfill
\includegraphics[width=0.47\textwidth,scale=0.2]{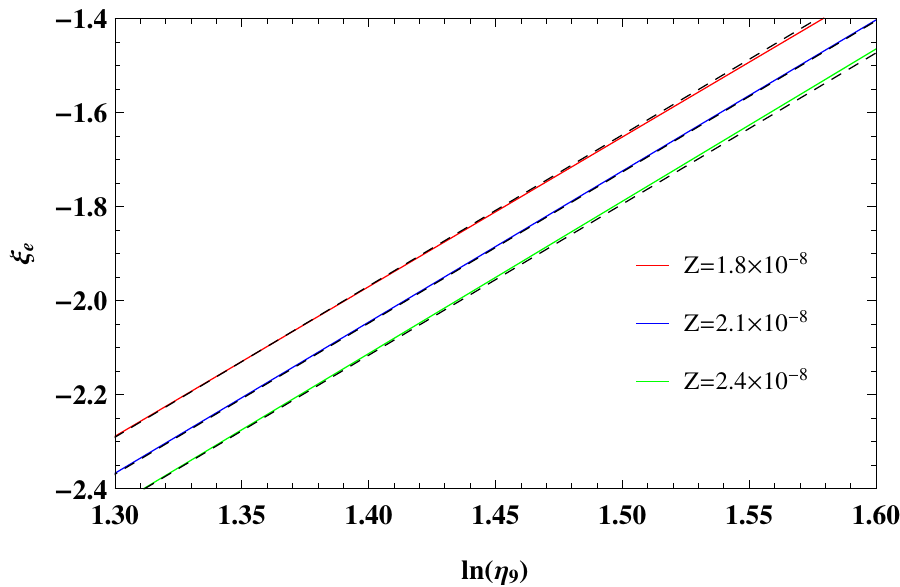}\hfill
\caption{Left Panel: LCN-predicted isoabundance curve for $^4$He in the $\ln(\eta_9)-\xi_e$ plane
for Y$_p$=0.22,0.24,0.26. Right Panel: Isoabundance curve for the total metallicity (Z) in the $\ln(\eta_9)-\xi_e$ plane
for Z=1.8$\times10^{-8}$, 2.1$\times10^{-8}$, 2.4$\times10^{-8}$. Solid lines are LCN predicted abundances using numerical code and dashed lines are the
corresponding rough analytic fits.}
\label{fit1}
\end{figure*}
\begin{figure*}[t!]
\centering
\includegraphics[width=0.47\textwidth,scale=0.2]{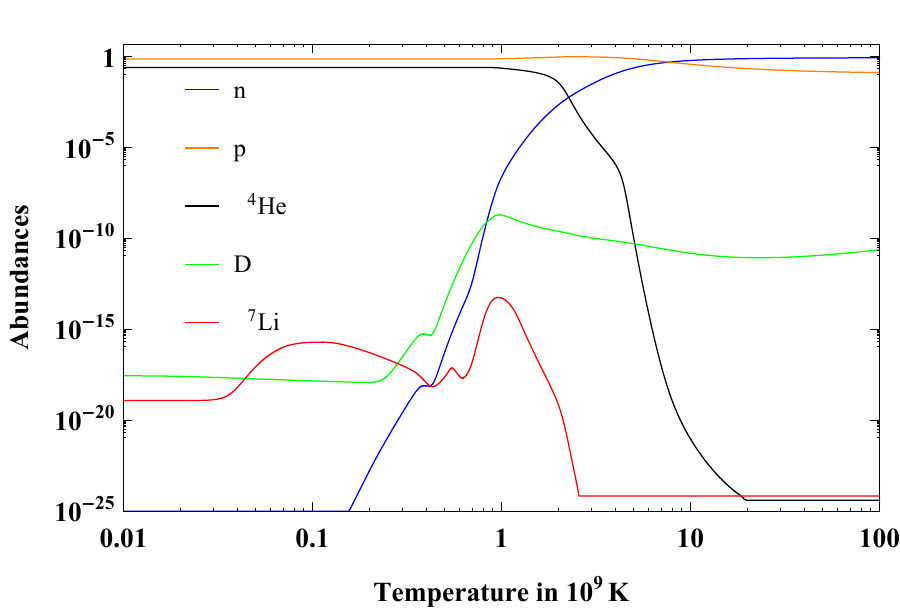}\hfill
\includegraphics[width=0.47\textwidth,scale=0.2]{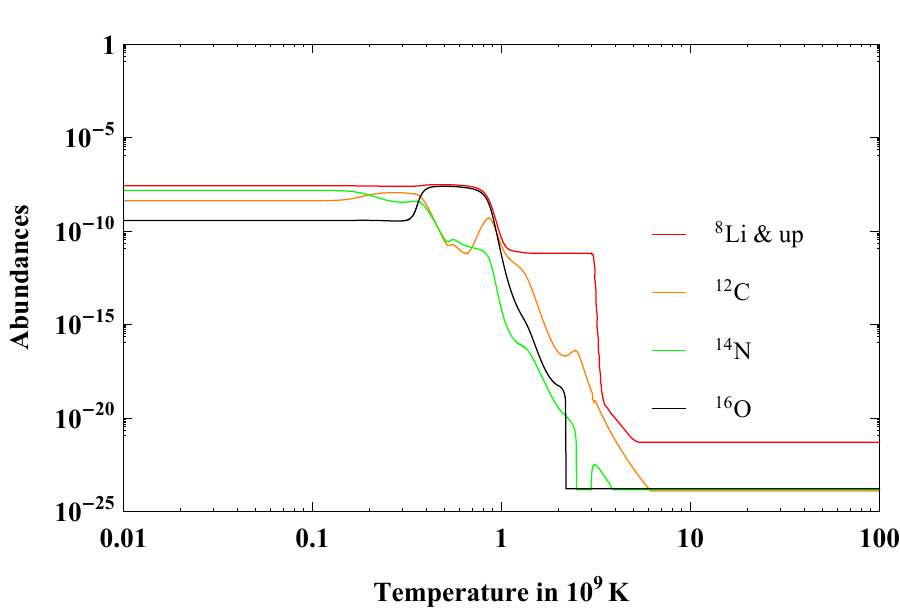}\hfill
\caption{Left Panel: Evolution of light element abundances with temperature in units 10$^9$ K. Using baryon to photon ratio
$\eta_9$=3.927, and electron neutrino degeneracy parameter $\xi_e$=-2.165. Right Panel: Evolution
of abundances of $^{12}$C,$^{14}$N,$^{16}$O as well as of A$>$8 elements (which include the CNO abundance) with temperature in units 10$^9$ K using same set of parameters.}
\label{abun}
\end{figure*}
As seen in \fig{fit1}, this fit holds good for 0.22 $\lesssim$ Y$_P$ $\lesssim$ 0.28, corresponding to -2.5 $\lesssim \xi_e \lesssim$ 0.
The primordial estimates of $Y_P$ inferred from observations \cite{izotov1} are:
\bes
\label{yp}
Y_P=0.254\pm 0.003. \\
\eta_{Y}=2.4515\pm 0.0207
\ees

\begin{figure*}[h!]
\centering
\includegraphics[width=0.5\textwidth,scale=0.2]{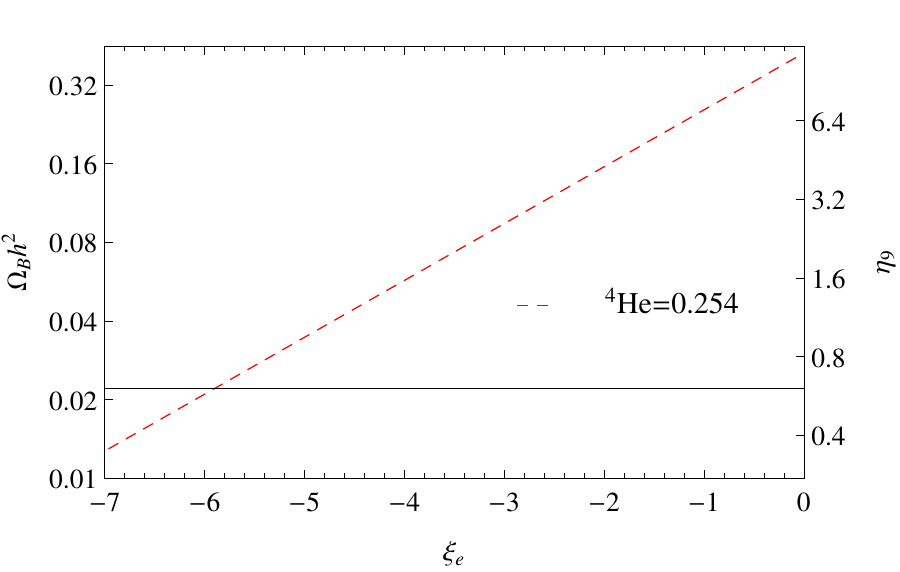}\hfill
\caption{The dashed (red) line describes the relation between $\eta_9$ and $\xi_e$
for a fixed observed $^4$He levels ($Y_P$). For $\eta_9$ = 0.6044, $\xi_e = -5.9$ }
\label{helomb}
\end{figure*}

We also used the code for $\eta_9$ varying between
0.3 and 10.0 and $\xi_e$ between 0 to -10.0. \fig{helomb} gives the relation 
between the two for a fixed predicted value of
$^4$He = 0.254. \\
\\
Constraints on the baryon entropy ratio $\eta_9$ = 0.6044 \cite{planck16} 
follow from WMAP and PLANCK data using the standard model as a prior. For LCC
a dedicated analysis is needed to analyse the said data and the above value
of $\eta_9$ is meaningless. However, just as a matter of interest, one can read 
from \fig{helomb} that $\xi_e$ = -5.9 is consistent with $\eta_9$ = 0.6044. \\
 \\

Similarly, an empirical fit of the total metallicity (Z) produced is:
\be
\ln(Z)=-17.6883\pm(.0074)+6.2832(\eta_{Z}-0.6637h-1.5461)\\
\ee
Where $\eta_Z$ is defined as:
\be
\label{etaz}
\eta_Z=\ln(\eta_9) - 0.3112\xi_e\hspace{1.2cm}
\ee
The minimum metallicity which is required for the
fragmentation and cooling process of prestellar gas clouds that
leads to the formation of lower mass PopII stars
is taken from \cite{sch2006,omu2005} as
\be
\label{z}
Z =Z_{cr} \equiv 10^{-6}Z_\odot\\
\ee
and from \cite{chap07} we have the solar metallicity Z$_\odot$ to be in the range
\be
0.0187 \leq Z_\odot \leq 0.0239
\ee
For calculation of the parameters, we consider the average value
\be
Z_\odot=0.0213.
\ee
This gives
\be
\eta_{Z}=2.0407\pm0.0251
\ee
\eqs{etay} and \eqs{etaz}  can be written as
\bes
\ln(\eta_9)=2.6405\eta_Z-1.6405\eta_Y\\
\xi_e = 5.2715(\eta_Z-\eta_Y)\hspace{.8cm}
\ees
Using constraints of $\eta_Y$ and $\eta_Z$ from \eqs{yp} and \eqs{z}, $\{\eta_9,\xi_e\}$, we get
\bes
\eta_9 = 3.927\pm 0.292\\
\xi_e= -2.165\pm 0.171
\ees
These constraints on $\{\eta_9,\xi_e\}$ in turn constrain $\{\Omega_B,L_e\}$.
\bes
\Omega_B = 0.263\pm 0.026\hspace{.2cm}\\
L_e =-0.795\pm 0.100
\ees
Assuming, $\xi_e=\xi_\mu=\xi_\tau$, gives a lower bound on $\eta_L$. The upper bound
 on $\eta_L$ is obtained from $\xi_\mu=\xi_\tau$=0.0 with only $\xi_e$ non vanishing. This gives
 \be
 -2.685 \lesssim\eta_L \lesssim -0.695
 \ee
These estimates of $\Omega_B$ saturate the dynamic mass estimates obtained from measured
velocity dispersion of stars in galaxy and of galaxies in clusters\cite{om}. This excludes any need of dark matter to account for these observed velocity dispersions. Thus a linear coasting universe with baryonic matter alone, having $\Omega_B = 0.263\pm 0.026$ and with a large Lepton asymmetry, would be concordant with observed values of $^4$He and the minimum metallicity. \\

From the above values of $\{\eta_9,\xi_e\}$, as a by product, one gets a higher amount of carbon($^{12}$C),
 nitrogen($^{14}$N) and oxygen($^{16}$O) in comparison with corresponding levels produced in SBBN.
\bes
\mathrm{X(^{12}C)}&\sim& 10^{-9}\nn\\
\mathrm{X(^{14}N)}&\sim& 10^{-8}\nn\\
\mathrm{X(^{16}O)}&\sim& 10^{-10}
\ees
These values of X(CNO) $\sim$ $\mathcal{O}(10^{-8})$ are high enough to match the observed metallicity in old low metallicity Type II stellar environments and also high enough to sustain a CNO cycle in a massive star \cite{cno}.
In SBBN, such metal enrichment requires an early generation of high mass PopIII stars having masses in the range: 10$^2$ -10$^5$ M$_{\odot}$. These have not been sighted to date - casting a serious doubt on their very existence.
However as we have seen above, nucleosynthesis in a linearly coasting universe eliminates the requirement of hypothetical Pop-III stars by producing the minimum metallicity levels (Z$_{cr}$=10$^{-6}$Z$_\odot$) in the early universe. \\

The problem with LCN are a rather low residual levels of deuterium($Y_D$) and lithium(Y$_{^{7}Li}$) that are produced in the early universe:
\be
Y_D\sim 10^{-18}; ~~~~ Y_{^7{Li}}\sim 10^{-19}
\ee
One might have expected that for $\Omega_B$ values an order of magnitude greater than in SBBN, copious amounts of $^7$Be would be produced and thereafter, by capturing an electron, yield a large amount of $^7$Li. However, the difference in the LCN scenario is that the relevant temperatures are held for time periods some eight orders of magnitude longer than in the SBBN. This ensures the destruction of $^7$Be as well as $^7$Li into the heavier nuclei channels over time period much smaller than the age of the universe at these temperatures in LCN. Thus by the time nucleosynthesis freezes, the residual levels of both $^7$Li as well as $^7$Be are abysmally low. \\

The only way out is to account for the production of D and $^7$Li to acceptable
levels in the later history of the universe. The plausibility of production of
both these elements to observed levels by spallation of high energy particles 
much later in the history of the universe is described in the next section \cite{gset}.\\ 
\\
\section{D and $^7$Li issues}
\begin{figure*}[h!]
\centering
\includegraphics[width=0.5\textwidth,scale=0.2]{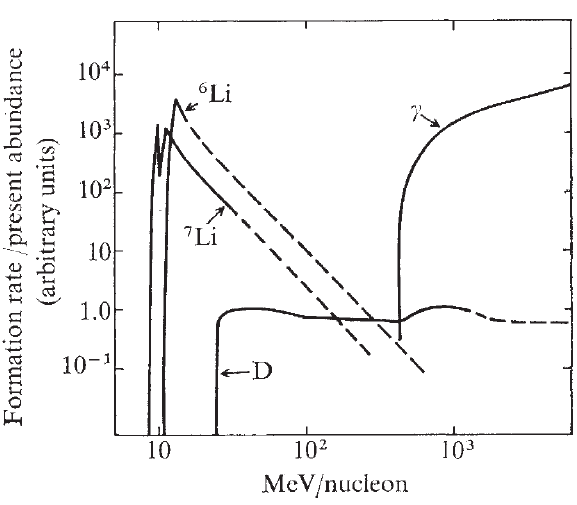}\hfill
\caption{Rates at which abundances approach the present values. Figure source: Ref. \cite{eps}}
\label{eps2}
\end{figure*}

One can only hope to reconcile with the low residual values of D and $^7$Li 
in early universe LCN, by some other means for their production to 
observed levels in the later history of the universe. 
We propose spallation mechanism in environments of incipient Pop-II stars 
as holding promise \cite{gset}. We revisit the results of Epstein 
et al \cite{eps} who clearly demonstrated that producing D to the observed 
levels by a spallation mechanism is not a problem at all. The problem 
is only the collateral over - production of $^7$Li levels.  
Their article considered spalling of accelerated beams of 
charged protons and alpha particles, with a relative composition of 
$\approx 25 \%$  of alpha particles by weight, over 
a cloud of type-I stellar composition. The results are summarized in the 
Fig.1 of their article reproduced in \fig{eps2}.
It can be read off from this figure that the production of D by a beam of protons
and alpha particles, with energy upto around 500 MeV/nucleon, spalling over an 
ambient cloud of similar composition, can produce D over the energy band 
$\approx$ 20 to 500 MeV/nucleon. However, for the energy band 10 MeV to 300 
MeV/nucleon, there is an unacceptable over production of $^7$Li. After such an 
over production of $^7$Li, its later destruction would completely destroy D. The 
only possibility is that of the spalling beam having a narrow energy band of 
around 400 MeV/nucleon. This would be physically hard to 
justify. Spallation mechanism was thus concluded as a ``no-go''.\\
\\
At the time Epstein et al wrote their article, models of star formation from 
protostellar clouds were still in their infancy. Essential aspects of evolution 
of a collapsing cloud to form a low mass Pop II star is now believed to be well 
understood \cite{feig,hart}. A protostar emerges from the collapse of a molecular 
cloud core surrounded by a high angular momentum circumstellar accretion disk. 
Such a star slowly contracts while magnetic fields play an important role in
regulating the collapse of a small fraction of the material of the disk into the
core while transfering the angular momentum of the disk to collimated outflows 
of the substantial major fraction of the disk material. Empirical studies on star
forming regions
over the last 25 years has provided ample evidence of beams of MeV particles 
produced by violent magneto - hydrodynamic reconnection events. Such flaring
is ubiquitous in all star forming regions. These are 
similar to solar magnetic flaring but elevated in intensity by factors of 10 to 
10$^6$ times 
the levels seen in the contemporary sun, besides being upto 100 times more 
frequent. Accounting for incipient sun's flaring in integrated effects of 
particle irradiation in the meteoritic record has assumed the status of an 
industry. Protons are the primary components of particle beams ejected from the sun
in gradual flares, while $^4$He are suppressed by factors of 10 in rapid flares 
up to 10$^3$ in gradual flares \cite{torsti,tere}. Models of the young sun 
visualize it as a much larger protostar with a cooler surface temperature and 
with a much higher level of magnetic activity in comparison to the  
contemporary sun. It is reasonable to assume that magnetic reconnection events
would lead to abundant release of MeV nucleii and strong shocks that would 
propagate into the circumstellar material. Considerable evidence for such 
processes has been found in the meteoric records. While it would be fair to say
that the hydrodynamic paradigm for understanding the earliest stages of stellar
evolution is still not complete, it would be reasonable to conjecture that 
features of collapse of a central core and its subsequent growth from accreting 
material would also hold for a low metallicity Pop II protostar. Strong magnetic 
fields would provide a link between the central star, its circumstellar envelope
and the accreting disk. Ejection of beams of ions from the surface of such
incipient central core would have extremely suppressed levels of $^4$He. Such a 
suppression is naturally expected as ionized particles are picked up from an 
environment that is cold enough to suppress ionized $^4$He in comparison to 
ionized hydrogen. Ionized helium to hydrogen ratio in a cool sunspot temperature
$\approx$ 3000 K can be calculated by the Saha ionization formula and the 
ionization energies of hydrogen and helium. This turns out to be 
$\approx exp(-40)$ and increases rapidly with temperature. Any electrodynamic 
process that beams out charged particles from this environment would yield beams 
predominantly consisting of protons and deficient in alpha particles. 
The temperature of the incipient core surface can be tweaked to produce 
any desired suppression of alpha particles.\\
\\
With such a natural mechanism for producing beams of particles deficient in 
$^4$He, 
the ``no-go'' concern of \cite{eps} gets effectively circumvented. In an incipient
Pop II environment, with levels of alpha particles suppressed by some three orders
of magnitude, $^7$Li production by $\alpha$ spalling over $\alpha$ would be 
naturally suppressed. Further, $^7$Li production by protons spalling over 
metals (CNO etc.) would be suppressed due to low metallicity in Pop II 
environments in comparison to the Pop I environment considered in \cite{eps}.
Thus there would be no problem of D production by spallation of beams of protons 
with energies less than some 500 MeV/nucleon.\\
\\
This would just leave $^7$Li production to be accounted for - as in SBBN. 
Referring back to
Fig. 1 of \cite{eps}, it is clear that low intensity cosmic rays with energies 
less than 20 Mev/nucleon would produce $^7$Li without disturbing D levels 
already produced by spallation. 

\section{Discussion and Conclusion}
Nucleosynthesis in a linearly coasting universe has characteristic features arising due to the slow expansion rate
of the universe at high temperatures. In previous studies it had been shown that production of concordant levels of $^4$He requires a baryon density ($^c\Omega_B \sim 0.70$): some three times the total dynamic mass estimates that follow from velocities of stars in galaxies and from the virial speeds of galaxies in clusters. $^4$He production levels are very sensitive to baryon density - rapidly vanishing for baryon density less than $^c\Omega_B$, and rapidly increasing to 100\% for higher values. One could in principle lower the average baryon density requirement in such a model by having an inhomogeneous distribution of baryons. The $^4$He levels are also much more sensitive to the nucleosynthesis rates due to the slow evolution in LCN than it is in SBBN. The reason is that in SBBN, the amount of $^4$He is completely determined by the $n/p$ ratio when deuterium burning becomes efficient. In LCN, $^4$He starts forming when the $n/p$ ratio is small but is still following its equilibrium value. This falls rapidly with temperature below 10$^{10}$K. With $n/p$ ratio small, the rate of $^4$He formation is small and dependent on the nuclear reaction rates, however, over the long time period at ones disposal at these temperatures, the abundance of $^4$He accumulates.\\

In this article we have explored the effect of a non-vanishing neutrino degeneracy parameter. In SBBN a non-vanishing degeneracy parameter has been invoked to get around the incompatibility of observationally inferred abundances of $^4$He. However, this does not auger well with the $^7$Li levels produced in SBBN \cite{steg07,steg08,cburt08}.\\

In LCN, $\xi_e$ determines
the initial $n/p$ ratio on the one hand and modifies the weak interaction rates that depend upon
number density of $\nu_e$ and $\bar\nu_e$. More positive the $\xi_e$, the lower is the neutron abundance and lower is the production of $^4$He. Similarly, more negative the $\xi_e$,
the larger is the neutron abundance and large the production of $^4$He. Using the present observational constraints on $^4$He, Y$_P$=0.254$\pm$0.003 and the minimum metallicity (Z) required for the fragmentation and cooling process in collapsing pre-stellar
gas clouds, Z=10$^{-6}$Z$_\odot$ (0.0187 $\leq$ Z$_\odot$ $\leq$ 0.0239), we put constraints on baryon to photon
ratio, the electron neutrino degeneracy parameters and the lepton asymmetry: $\eta_9$ = 3.927$\pm$0.292 (which is equivalent to $\Omega_B$ =0.263$\pm$0.026),
$\xi_e$ = -2.165$\pm$0.171 (equivalent to L$_e$=-0.795$\pm$0.100). Assuming $\xi_e=\xi_\mu=\xi_\tau$, we get a lower bounds on $\eta_L\geq$-2.685. On the other hand if  we assume $\xi_\mu=\xi_\tau=0.0$ and only $\xi_e$ non-vanishing, we get $\eta_L\le$-0.695. With such a constraint on $\Omega_B$,  the dynamic mass estimates can be saturated by baryonic mass density alone. Another useful by product of the slow expansion rate is the production of significant X(CNO)$\sim 10^{-8}$. These abundances are sufficient to trigger and sustain the CNO cycle in low metallicity stars. The overall metallicity produced is also sufficient for efficient cooling in collapsing star forming clouds. \\
\\

It is interesting to compare the production of $^7$Li and $^7$Be in LCN and SBBN.  There are two aspects of the issue of generating Lithium. 
Firstly, it is the production in the early hot stage and then any
later production by spallation in the later history of the Universe. 
In the Early hot stage, in the SBBN scenario, 
lithium is determined by competing processes of production, by $^4$He + $^3$H $\longrightarrow$ $^7$Li + $\gamma$,
and its destruction by $^7$Li + p $\longrightarrow$ $^4$He + $^4$He. With growing $\eta$, the destruction by the later
 reaction depresses the amount of $^7$Li produced. However, for $\eta$ $>$ 3$\times$10$^{-10}$ or 
so, Beryllium starts getting copiously produced by $^4$He + $^3$He $\longrightarrow$ $^7$Be + $\gamma$. $^7$Be 
is then converted into $^7$Li by electron capture: $^7$Be + e$^-$ $\longrightarrow$ $^7$Li + $\gamma$, 
with the destruction channel the same as for low $\eta$. 
The extra channel leads to an increase in $^7$Li with an increase of the baryon to photon ratio.
 This entire scenario is completely different in a cosmology with a Linearly evolving scale factor. 
In LCN, weak interactions are in equilibrium till much lower temperatures -
leading to a very low n/p ratio. The D, $^3$H, and $^3$He  levels are considerably (and abysmally) low throughout. 
$^4$He being stable, its levels build up slowly due to the long time (hundreds of years) that the universe is held 
at these temperatures. For the same reason, whatever $^7$Li is produced, by either channel, is destroyed by the same 
destruction process as in SBB: $^7$Li + p $\longrightarrow$ $^4$He + $^4$He. In LCN, thus,
in the early universe nucleosynthesis epoch, we can get the desired amount of 
$^4$He and significant metallicity [CNO], with virtually negligible amounts of other light elements. As shown in the previous section, D and $^7$Li can in principle
be produced later in the history of the universe.  \\
\\

An over all comparison of a Linear Coasting model with the standard ``Big-Bang 
Model'', 
would be quite premature at this stage. The effort that has gone in the study of 
concordance of SBB (the $\Lambda_{CDM}$ model) with observations is quite a 
commendable task and no such comprehensive consolidated effort is in place for 
a Linear Coasting model. Constraints on the baryon density $\Omega_B$
that have been obtained from microwave background anisoropy measurements 
by the WMAP and Planck satellite experiments use the Standard Big Bang model 
(the $\Lambda_{CDM}$ model) as a prior. 
The concordance of D levels, that follow from this $\Omega_B$, with the 
measurements of Cooke et al \cite{cooke} in Quasar Absoption
Systems is enviably remarkable. For our purpose, taking note that these systems 
have signatures of significant 
metallicity, it is clear that the absorption system has to follow a metal 
enriching phase. All we hope is that if the process of such enrichment involves 
star formation, the flaring events accompanying 
such incipient Pop - II stellar environments could produce  D to the levels 
observed by Cooke et al even in LCC.\\ 

The above remarkable concordance of the inferred baryon density in SBB from SBBN 
and CMB anisotropy measurements must be matched in
a linear coasting model for the corresponding 
$\Omega_B$ deduced therein before it stakes a claim to be 
a viable alternative. We are continuing our effort in this direction. The overall
strategy is to study CMB anosotropy with the LCC prior and determine constraints 
on $\Omega_B$ that follow from it. Using this value of $\Omega_B$, the observed
$^4$He levels would determine the $\xi_e$ and the metallicity as described in this 
article. Though our work on CMB anisotropy is still in progress and as yet 
incomplete, we conclude by mentioning some encouraging results.  
In SBB, the growth of linear density perturbation in the radiation dominated epoch 
can be expressed, in conformal time $\eta$, as a superposition of two modes - one 
that blows up at $t = \eta = 0$, and the other that remains bounded. Requiring the 
modes to be bounded at the initial time, one evolves such perturbations till the last 
scattering surface to get a frozen pattern of phase correlated modes at the epoch 
of decopuling. In a linear coasting cosmology, one has to find some other 
explanation for phase correlation. This is because $t = 0$ corresponds to the 
conformal time $\eta = - \infty$ in LCC and a superposition of both the modes 
at $t = 0$ is not ruled out as they are both bounded in LCC. Assuming that sound 
waves are generated at a finite epoch in the past - say at the epoch of the QGP 
phase transition, these sound waves can once again be propagated to the last 
scattering surface. There are some remarkable coincidences: (a) the Hubble scale at 
the last scattering surface in LCC subtends an angle of the same order corresponding
to the observed first peak in the CMB anisotropy, and (b) The peak locations and 
intervals of the first three peaks match with the results in SBB. (c) An account 
of growth of structure in LCC does not to pose much problem \cite{savthesis}, \\
\\
Unfortunately, other than these two coincidences, which no doubt are quite
remarkable in view of the enormous difference of LCC in comparison to SBB, we have
not yet been able to match the CMB anisotropy spectrum in LCC with observations 
anywhere near the precission done in SBB with the $\Lambda_{CDM}$ prior. Thus, while
we have not given up yet, with an account of CMB anisotropy still 
wanting, it would be fair to say that as on date it is the CMB anisotropy that 
rules out a linear coasting cosmology.\\ 
\\
The purpose of the present study is restricted to the determination of 
constraints on nucleosynthesis parameters in a linearly coasting cosmology - 
awaiting our ongoing effort of seeking concordance of CMB anisotropy.

 \section*{Acknowledgements}
This work is financially supported by SRF grant(09/045(0934)/2010-EMR) from CSIR, India. We would like to thank the referee for useful comments.

\appendix

\section{Analytical expression for neutron and proton destruction rates}
\label{apn}

The rates for the reactions given in \eqs{reaction}are
\cite{bg1976,weinberg}

\bes
\label{1}
\lambda_p &=& K\left(\int_{1}^{\infty}\frac{x(x^2-1)^{1/2}(x+q)^2\mathrm{d}x}{(1+e^{-xz})(1+e^{(x+\beta)z_\nu})}+
\int_{1}^{\infty}\frac{F(x)x(x^2-1)^{1/2}(x-q)^2\mathrm{d}x}{(1+e^{xz})(1+e^{-(x-\beta)z_\nu})}\right)
\ees
\bes
\label{2}
\lambda_n &=& K\left(\int_{1}^{\infty}\frac{x(x^2-1)^{1/2}(x+q)^2\mathrm{d}x}{(1+e^{xz})(1+e^{-(x+\beta)z_\nu})}+
\int_{1}^{\infty}\frac{F(x)x(x^2-1)^{1/2}(x-q)^2\mathrm{d}x}{(1+e^{-xz})(1+e^{(x-\beta)z_\nu})}\right)
\ees

Where, q = $\frac{m_n-m_p}{m_e}$ $\simeq$ 2.531\\
$m_n,m_p,m_e$ are mass of neutron, proton and electron respectively.\\
z = m$_e$c$^2$/kT$_\gamma\simeq$ 5.930/T\\
z$_\nu$ = m$_e$c$^2$/kT$_\nu \simeq$5.930/T$_\nu$\\
$\beta$ = $\xi_e$/z$_\nu$ + q\\
K=6.79$\times$10$^{-4}$ sec$^{-1}$\\
The function F(x) is the fermi function. Throughout the relevant range in our case where it makes a significant contribution, its rough constant value is $\simeq$ .98 \\
For z large compared with unity, we have z$_\nu$ = z this gives:
$\lambda_n$ = e$^{\beta z}\lambda_p$\\
The neutron to proton ratio is given by the ratio of proton destruction rate to
neutron destruction rate ($\lambda_p/\lambda_n$) which implies the expression given in equation \eqs{np}.

\bibliographystyle{utcaps}
\providecommand{\href}[2]{#2}\begingroup\raggedright\endgroup
\end{document}